\documentclass[a4paper,11pt]{article}
\usepackage{jheppub}
\usepackage{mathrsfs}
\usepackage{amsfonts}
\usepackage{setspace}
\usepackage{cellspace}
\usepackage{amsmath,bm}
\usepackage[colorlinks=true,linkcolor=blue]{hyperref}
\usepackage{xcolor}
\usepackage{epsfig}
\usepackage{slashed}
\usepackage{caption}
\usepackage{hhline,multirow,tabularx}  % for nicer tables
\usepackage{dcolumn}    % align table columns on decimal point
\usepackage{url}        % for URL addresses
\usepackage{braket} 
\newcommand{\re}{{\rm e}}
\newcommand{\ri}{{\rm i}}
\newcommand{\rd}{{\rm d}}
\newcommand{\mq}{\mathfrak{q}}
\title{Analyticity, asymptotics and natural boundary for a one-point function of the finite-volume critical Ising chain}
\author[a]{Yizhuang Liu}

\affiliation[a]{Institute of Theoretical Physics, Jagiellonian University, 30-348 Kraków, Poland}

\emailAdd{yizhuang.liu@uj.edu.pl}

\abstract {This note reports the following observation: the finite-volume expectation value of the spin operator (the one-point function) between the $\mathbb{Z}_2$-even and odd ground states in the critical periodic Ising chain, when continued as a complex-analytic function of the system length $N$ through the Borel resummation of its large-$N$ expansion, has a natural boundary of analyticity along the negative real axis. The singular behavior near the negative real axis, after an exponential map, is the same as that of a Lambert-type series
for the odd-divisor-squared sum near the unit circle $|z|=1$. The same divisor sum also governs the strengths of the Borel discontinuities of the one-point function's factorially-divergent large-$N$ asymptotics. We also report the all-order large-$N$ asymptotics of the leg function for the finite-volume spin-operator form factor, and the similarities to certain known quantities in the literature. 
} 
\date{\today}

\begin{document}
\maketitle
\flushbottom

Natural boundaries of analyticity are common in mathematical literature, but not that common in ground state properties of integrable lattice models, especially at infinite system sizes. The most famous example in this context is probably the Nickel's singularities of the Ising magnetic susceptibility~\cite{BernieNickel_1999,Orrick_2001}. This note will provide another example. We will show that there is a natural boundary for the standard Ising-chain's finite volume one-point function at the critical parameter~\cite{Liu:2026qxw}
\begin{align}
|\langle \Omega_R|\hat \sigma| \Omega_{NS}\rangle|^2=G\left(\frac{1}{2}\right)G\left(\frac{3}{2}\right)\left(\frac{2\pi}{N}\right)^{\frac{1}{4}} \re^{v\left(\frac{1}{N}\right)} \ . \label{eq:onepoint}
\end{align}
Here, $\hat \sigma$ is the spin-operator, the $|\Omega_{NS}\rangle$ and $|\Omega_R\rangle$ are the  {\it absolute ground states} in the $\mathbb{Z}_2=\pm 1$ sectors, and the $N$ is the system size. Equivalently, the same quantity can also be interpreted as the overlap between two absolute ground states, one for the periodic chain, and another for the anti-periodic chain. The factor $\left(\frac{2\pi}{N}\right)^{\frac{1}{4}}$ is of CFT origin, while the constant $G\left(\frac{1}{2}\right)G\left(\frac{3}{2}\right)$ enforces the ``simple fixed-point scaling''~\cite{McCoy:2000gw} for the finite-volume scaling limit. 

The function $v\left(\frac{1}{N}\right)$ is the focus of this note. It measures the power-corrections to the CFT limit and can be simplified to a single integral~\cite{Liu:2026qxw}
\begin{align}
v\left(\frac{1}{N}\right)=-\int_0^{\infty} \rd t \frac{\tanh^2 \frac{t}{4N} }{2t(\re^t+1)} \ .  \label{eq:defv}
\end{align}
In~\cite{Liu:2026qxw}, this is only proven for $N\in2 \mathbb{Z}_{\ge 2}$, but it is also valid for odd system sizes. As we will see latter, the large $N$ expansion of $v \left(\frac{1}{N}\right)$ is factorially divergent, but it is Borel summable to itself and defines an analytic function in $N$.  Meanwhile, the $N \rightarrow 0$ expansion of $v$ is also factorially divergent. This is already rather different from functions such as $\psi\left(1+N\right)$, $\ln \Gamma \left(1+N\right)$ and $\ln G(1+N)$. As we will show,  there is a natural boundary when analytic continued to the negative real axis, starting from the positive real axis. And the limiting behavior near the boundary is controlled by a non-modular Lambert series for a divisors-squared sum, which also governs the irregularities in the large-$N$ factorial asymptotics and the strengths of its   Borel discontinuities. 

After finishing the first version of the note, which consists essentially the first two sections, the author has observed, that similar asymptotics have appeared and been studied previously in other contexts~\cite{Pasquetti:2010bps,Rella:2022bwn,Gu:2023mgf}. In fact, the one-point function Eq.~(\ref{eq:onepoint}) itself can also be expressed in terms of known quantities in the literature~\cite{Periwal:1993yu,Tierz:2002jj,Marino:2004uf}. We will present the observed similarities in more details in Sec.~\ref{sec:compare}.

\section{Analytic continuation and behavior near the negative real axis}
Naively looking, Eq.~(\ref{eq:defv}) defines an even function in $N$, that is analytic when $N$ is away from the imaginary axis. When $N$ is imaginary, there are infinitely many poles along the integration path and the integral is not well defined.  As such, in the naive way of analytic continuation, there would be a branch cut along the imaginary axis. But this branch cut is not really there, because the left and right limits, although differ by a sign in the non-vanishing imaginary parts, are still regular and can be continued further to the other sides. We will start from the positive real $N$, and analytic continue to the left half plane.  Another version is just its mirror. When continued close enough to the negative real axis, the behavior starts to become bad enough and further analytic continuation is no longer possible. The main focus of this note is to show this in details and reveal the behavior near the negative real-axis.

\subsection{Analytic continuation and reflection formulas}

To perform the analytic continuation to the left half-plane, Eq.~(\ref{eq:defv})
is not very convenient. One can use the following Mellin-Barnes representation for $N>0$
\begin{align}
v\left(\frac{1}{N}\right)=2\int_{-2<{\rm Re}(u)<0}\frac{\rd u}{2\pi \ri}\left(2^u-4\right) \left(2^{u+1}-1\right)  \zeta (u-1) \zeta (-u) \frac{\pi}{u\sin \pi u}\left(\frac{1}{N}\right)^{-u} \ . \label{eq:Barnes}
\end{align}
The integration path is along the imaginary axis from ${\rm Re}(u)-\ri\infty$  to ${\rm Re}(u)+\ri\infty$. By shifting the contour to the left, one generates systematically the large $N$ asymptotics given latter in Eq.~(\ref{eq:factorialB}).  Eq.~(\ref{eq:Barnes}) also establishes the $v$ as an analytic function in the complex $N$-plane in the sectors $|{\rm Arg} (N)|<\pi(1-\delta)$ for any $0<\delta<1$. This is because the integrand decays exponentially in such sectors.
When $N$ approaches the negative real axis, however, the exponential decay speed $e^{-\pi |{\rm Im}(u)|}$ of the $\frac{1}{\sin \pi u }$ is canceled by the phase angle of the $N$, and the behavior of the integral is then dominated by the highly oscillating and non-predictable products of the two $\zeta$ functions. This already suggests that the regularity of the boundary limit to the negative real axis could be very low.

To demonstrate this explicitly,  it is more convenient to use the following reflection formula when $-\pi<{\rm Arg}(N)<\pi$
\begin{align}
v\left(-\frac{1}{N}\right)=-v\left(\frac{1}{N}\right)+\frac{1}{2}\sum_{n=0}^{\infty}(-1)^n(n+1)\ln \frac{n(n+2) \tan^2\frac{\pi(n+1)}{4N}}{(n+1)^2 \tan \frac{\pi n }{4N} \tan \frac{\pi(n+2)\pi}{4N}} \ . \label{eq:reflection1}
\end{align}
The $n=0$ term is understood as the $n\rightarrow 0$ limit.
The behavior near the negative real axis is then reduced to the infinite sum on the right hand side of Eq.~(\ref{eq:reflection1}). For notation simplicity, it is convenient to change the variable according to 
\begin{align}
\frac{1}{N} = 2(x+\ri y) \ , \  v\left(2(x+\ri y)\right) \equiv v(x,y) \ .
\end{align}
In terms of the $v(x,y)$, for $y>0$, the reflection formula can be simplified to 
\begin{align}
&v(-x-\ri y)=-v(x+\ri y)+\frac{1}{2}\sum_{n=1}^{\infty}(-1)^n (n+1)\ln \frac{n(n+2)}{(n+1)^2}\nonumber \\ 
&+\frac{1}{2} \ln \frac{4\tan^2 \frac{x+\ri y}{2}}{\pi(x+\ri y) \tan (x+\ri y)}+{\cal S}_r(x,y)+{\cal S}(x,y) \ . \label{eq:reflection2}
\end{align}
In the above, the regular part of the sum is 
\begin{align}
{\cal S}_r(x,y)=\sum_{k=0}^{\infty}\frac{1}{2k+1}\bigg(\frac{2}{w^{2k+1}}-\frac{1}{w^{4k+2}}\bigg) \ , \  \ w=\re^{-\ri \pi x}\re^{\pi y } \ , 
\end{align}
and the singular sum ${\cal S}$ is 
\begin{align}
{\cal S}(x,y)=-\sum_{k=0}^{\infty}\frac{1}{2k+1}\frac{4w^{2k+1}}{(w^{2k+1}+1)^2} \  ,  \ w=\re^{-\ri \pi x}\re^{\pi y }  \ . 
\end{align}
The $y<0$ case can be obtained from the above with $(x,y) \rightarrow (-x,-y)$. Notice the singular part itself is invariant under $(x,y) \rightarrow (-x,-y)$.
The regular sum can be performed exactly into logarithmic functions and there is no natural boundary: 
\begin{align}
{\cal S}_r(x,y)=\ln \frac{1+\re^{\ri\pi x}\re^{-\pi y}}{1-\re^{\ri\pi x}\re^{-\pi y}}-\frac{1}{2}\ln \frac{1+\re^{2\ri\pi x}\re^{-2\pi y}}{1-\re^{2\ri\pi x}\re^{-2\pi y}} \ . 
\end{align}
Further more, for both $y>0$ and $y<0$ one has the following 
\begin{align}
{\cal S}_r(x,y)+\frac{1}{2} \ln \frac{4\tan^2 \frac{x+\ri y}{2}}{\pi(x+\ri y) \tan (x+\ri y)}=\frac{1}{2}\ln \frac{4\ri {\rm sign}(y)}{\pi(x+\ri y)} \ ,
\end{align}
and one has the following sum
\begin{align}
-\frac{1}{2}\sum_{n=1}^{\infty}(-1)^n (n+1)\ln \frac{n(n+2)}{(n+1)^2}=\ln \bigg[2G^2\left(\frac{1}{2}\right)G^2\left(\frac{3}{2}\right)\bigg] \ . 
\end{align}
The above allow to further simplify the reflection formula for all $y\ne 0$ as
\begin{align}
v(-x-\ri y)+v(x+\ri y)+\ln \bigg[\pi^{\frac{1}{2}}G^2\left(\frac{1}{2}\right)G^2\left(\frac{3}{2}\right)\bigg]= \frac{1}{2}\ln \frac{\ri {\rm sign}(y)}{x+\ri y}+{\cal S}(x,y) \ . 
\end{align}
Clearly, the presence of ${\rm sign}(y)$ is not in contradiction to analyticity, as the analyticity domain of $v(-x-\ri y)+v(x+\ri y)$ contains two disconnected components, one is the upper half-plane, and another is the lower-half.

\subsection{Behavior near the natural boundary}

As such, the behavior near the negative real axis  reduces to the $y \rightarrow 0$ limit for the singular sum ${\cal S}(x,y)$.  We first consider the real part, which can be written as
\begin{align}
{\rm Re} ({\cal S})(x,y)=-\sum_{k=0}^{\infty} \frac{2}{2k+1} \frac{1+\cos (2k+1)\pi x \cosh (2k+1)\pi y }{(\cos (2k+1)\pi x+\cosh (2k+1)\pi y)^2} \ . \label{eq:realS}
\end{align}
As such, when $|y|>0$, the sum is absolutely convergent and there are no singularities. However, when $y \rightarrow 0$, the asymptotic behavior depends on the  number-theoretical properties of the boundary point $-x$.

We first consider the case when $x$ is a dyadic rational number of the form
\begin{align}
x=\frac{2q+1}{2^p } \ , \ p> 0 \ .
\end{align}
In this case, one has the following asymptotic behavior 
\begin{align}
{\rm Re}\big({\cal S}\big)\left(\frac{2q+1}{2^p },y\right)\bigg|_{p \ne 0, \ y\rightarrow 0}=\frac{2^p}{2} \ln |y| +{\cal O}(1) \ . \label{eq:asymdya}
\end{align}
This already implies the existence of a natural boundary, since dyadic numbers form a dense subset. The coefficient of the $\ln y$ can be made rigorous in the following way: one splits the sum at $k_0=[\frac{1}{2\pi y}]$, and subtracts from the sum in the $0\le k\le k_0$ region, the summand at $y=0$. This is legal, since one has the following inequality  
\begin{align}
{\rm min}_{0\le k\le 2^p-1} \bigg(1+\cos \frac{(2k+1)(2q+1)\pi}{2^p}\bigg) >0 \ , \label{eq:bounds}
\end{align}
when $p>0$. Given the above, it is not hard to show that the subtracted sum in the region $0\le k\le k_0$, and the original sum in the region $k>k_0$ are all bounded from the above by expressions that are ${\cal O}(1)$ when $y \rightarrow 0^+$. As such, one has 
\begin{align}
{\rm Re}\big({\cal S}\big)\left(\frac{2q+1}{2^p },y\right)=-\sum_{k=0}^{[\frac{1}{2\pi |y|}]} \frac{2}{2k+1} \frac{1}{1+\cos \frac{(2k+1)(2q+1)\pi}{2^p}}+{\cal O}(1) \ .  \label{eq:asym1}
\end{align}
Namely, the leading small-$y$ singularity is encoded by the summation at $y=0$ and truncated up to ${\cal O}(\frac{1}{|y|})$. Now, one sums for different $k \mod{2^{p}}$, and 
use the following summation formula
\begin{align}
\sum_{k=0}^{2^p-1} \frac{1}{1+\cos \frac{(2k+1)\pi}{2^p}}=2^{2p-1} \ , \label{eq:sum1}
\end{align}
to extract the coefficient of the logarithmic term. The case of other rational numbers with factors of $2$ can be established similarly. For more details of the arguments above, see Sec.~\ref{sec:numberdetails}

We then consider the case where there are no factors of $2$. In this situation the divergences as $y\rightarrow 0^+$ are much faster and is of the order $y^{-2}$. This is because in this case, the bound Eq.~(\ref{eq:bounds}) is invalid, and the leading singularities are determined by the small denominators for which $\cos (2k+1)\pi x$ equals $-1$. Now, for
\begin{align}
x=\frac{2q_1+1}{2q_2+1} \ , 
\end{align}
in which the $2q_1+1$ and $2q_2+1$ are co-primes, the $\cos (2k+1)\pi x=-1$ when 
\begin{align}
2k+1=(2l+1)(2q_2+1) \ , \ l\in\mathbb{Z}_{\ge 0} \ . 
\end{align}
In the case, the leading singularity can be extracted by collecting all such terms
\begin{align}
\frac{2}{2q_2+1}\sum_{l=0}^{\infty} \frac{1}{2l+1}\frac{1}{\cosh \pi(2l+1)(2q_2+1)y-1} \rightarrow \frac{7\zeta_3}{2\pi^2} \frac{1}{(2q_2+1)^3} \frac{1}{y^2} \  ,
\end{align}
which is dominated by the leading small-$y$ expansion of $\cosh \pi(2l+1)(2q_2+1)y-1$ in the denominator. To summarize, for rational numbers, one has the following leading singularities
\begin{align}
&x=\frac{2q_1+1}{2q_2+1}:  \ \frac{7\zeta_3}{2\pi^2}\frac{1}{(2q_2+1)^3} \frac{1}{y^2} \ , \label{eq:asymodd} \\
&x=\frac{2q_1+1}{2^p(2q_2+1)}:   \ \frac{2^p(2q_2+1)}{2} \ln |y| \ ,  \ p>0 \ , \label{eq:asymeven1}\\ 
&x=\frac{2^p(2q_1+1)}{(2q_2+1)} :  \ \frac{2q_2+1}{2} \ln |y|  \ ,  \ p>0 \ . \label{eq:asymeven2}
\end{align}
 Namely, when there are no factors of $2$, the singularity is quadratic. When there are factors of $2$, the speed is logarithmic, and the coefficient is half of the smallest denominator. 
 
 For irrational numbers, the  $\cos (2k+1)\pi x$ can become very close to $-1$, and the leading small-$y$ singularity is determined by the accumulation of such small denominators before the $\cosh (2k+1)\pi y$ becomes too large. However, we know that $\cos (2k+1)\pi x$ becomes close to $-1$ when $x$ becomes close to $\frac{2q+1}{2k+1}$. As such, the questions then reduces to ask when and how good a irrational number $x$ can be approximated by a ratio of two odd integers, up to a given truncation level of the denominator. It might be hard to find a general formula for the leading singularity for generic irrational $x$, but in any case, the $y\rightarrow 0$ singularity always satisfies the following 
 power-law bound
\begin{align}
|{\cal S}(x,y)|\le \frac{1}{2 y^2}  \  , 
\end{align}
 which we will show at the end of this section, after establishing the connection to the function $\sigma_{-2}^o(n)$.  The presence of a natural boundary in the analytic continuation in the system size might be related to certain lattice-level properties of the Ising chain, that also depend on number-theoretical properties of the system size $N$.

We also comment on the imaginary part, but the discussions will be less detailed than the real part, as the imaginary part is less singular.  Again, we only need to consider the singular sum ${\cal S}$. The imaginary part can be extracted as
\begin{align}
{\rm Im}({\cal S})(x,y)=-\sum_{k=0}^{\infty}\frac{2}{2k+1}\frac{\sin (2k+1)\pi x \sinh (2k+1)\pi y}{(\cos (2k+1)\pi x+\cosh (2k+1)\pi y)^2} \ . 
\end{align}
The major difference to the real part is, when $\cos (2k+1)\pi x=-1$, the $\sin (2k+1)\pi x$ is forced to vanish. For rational numbers, this removes the quadratic divergences in case without factors of $2$, and as $y\rightarrow 0$ the ${\rm Im}{\cal }({\cal S})(x,y)$ all approach to $0$ linearly. But the speeds are different at different $x$. On the other hand, the regular part ${\cal S}_r(x,y)$ as $y\rightarrow 0^+$ is non-vanishing for generic $x$, thus for rational numbers the limiting behavior of the imaginary parts are also dominated by the regular part. However, for irrational $x$, the singular part ${\rm Im}({\cal S})(x,y)$ still diverges when $y \rightarrow 0$.    
\begin{figure}[htbp]
    \centering
    \includegraphics[height=9.0cm]{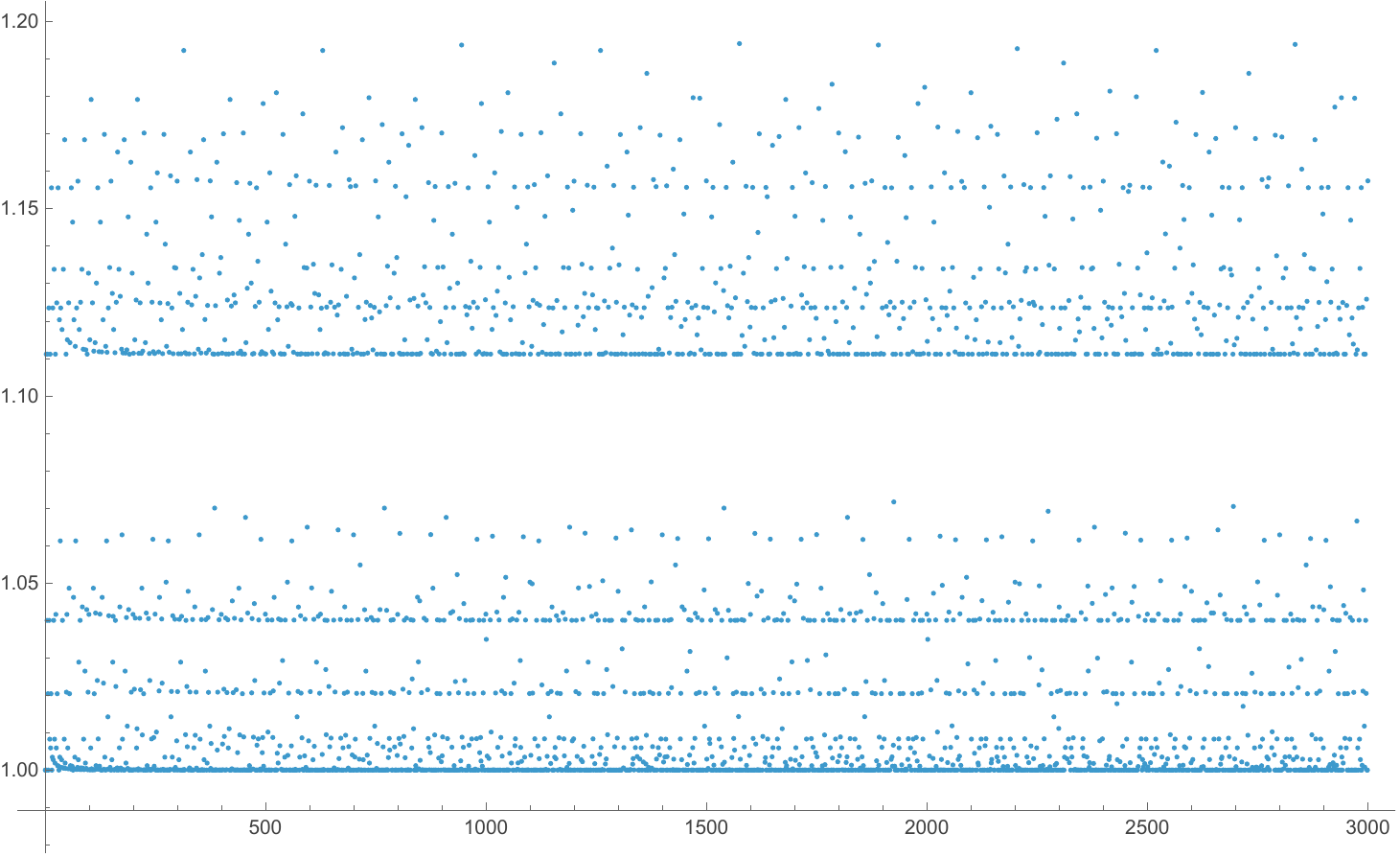}
    \includegraphics[height=9.0cm]{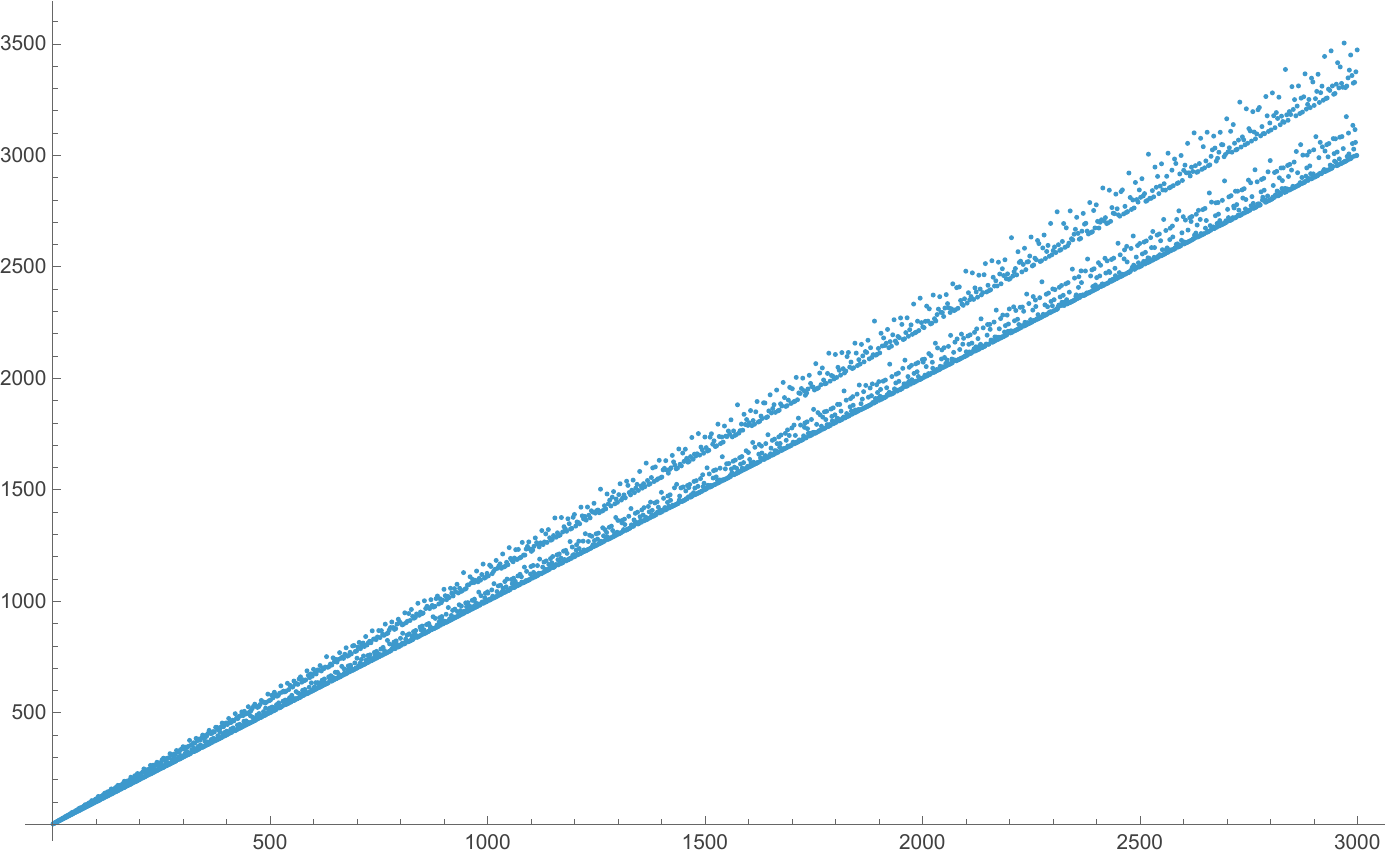}
    \caption{Plots of the functions $\sigma_{-2}^o(n)$ (upper) and $n\sigma_{-2}^o(n)$ (lower) for $1\le n\le 3000$. }
    \label{fig:rn}
\end{figure}

\subsection{Divisor-square sums and Lambert series}
We have seen that the singular behavior near the negative real axis is encoded by the singular sum 
\begin{align}
{\cal S}(z)=-\sum_{k=0}^{\infty}\frac{4}{2k+1}\frac{z^{2k+1}}{(z^{2k+1}+1)^2} \ ,  \ z= \re^{\ri\pi x}\re^{-\pi y} \ , 
\end{align}
when $y \rightarrow 0^+$. This suggest to introduce the following analytic function ${\cal S}(z)$ defined by the Taylor expansion in the region $|z|<1$
\begin{align}
{\cal S}(z)=-4\sum_{l=0}^{\infty}\sum_{k=0}^{\infty}(-1)^l\frac{l+1}{(2k+1)}z^{(l+1)(2k+1)}=4\sum_{n=1}^{\infty}(-1)^n z^n n\sum_{\substack{l |n \\ l\ge 1}}\frac{1-(-1)^l}{2l^2} \ . 
\end{align}
The $y\rightarrow 0^+$ limit is then mapped to $|z|\rightarrow 1^-$. The reason for the singular behavior near the unit circle is then made clear: it is due to the irregularities in the divisor sum
\begin{align}
\sigma_{-2}^o(n)=\sum_{\substack{l |n \\ l\ge 1}}\frac{1-(-1)^l} {2l^2} \ , \label{eq:defrn}
\end{align}
for which ${\cal S}(-z)$ is essentially its generating function
\begin{align}
{\cal S}(-z)=4\sum_{n=1}^{\infty}n\sigma_{-2}^o(n) z^n \ .
\end{align}
For all $n\ge 1$, $\sigma_{-2}^o(n)$ is bound from below by $1$, and from the above by $\frac{3}{4}\zeta_2$ 
\begin{align}
1\le \sigma_{-2}^o(n)<\frac{\pi^2}{8}=\sum_{k=0}^{\infty} \frac{1}{(2k+1)^2}\ .
\end{align}
This implies the following power-law bound
\begin{align}
|{\cal S}(z)|\le 4\times \frac{\pi^2}{8}\frac{|z|}{(1-|z|)^2} \le \frac{1}{2y^2} \ ,
\end{align}
which is indeed satisfied in all the known cases Eq.~(\ref{eq:asymodd}) to Eq.~(\ref{eq:asymeven2}). Due to the power-law bound, the boundary limit still exists in the distributional sense. We have included two plots in the Fig.~\ref{fig:rn}, to depict the fluctuations in $\sigma_{-2}^o(n)$ and $n\sigma_{-2}^o(n)$. 

The appearance of the divisor sum suggests that ${\cal S}(z)$ can be further simplified to the form of Lambert series~\cite{schmidt2020cataloginterestingusefullambert}. We introduce the notations $\sigma_k(n)=\sum_{p|n}p^k$,  $\sigma^e_k(n)=\frac{1}{2}\sum_{p|n}(1+(-1)^p)p^k$, $\sigma^o_k(n)=\frac{1}{2}\sum_{p|n}(1-(-1)^p)p^k$ for the full, even and odd divisor sums. Then, one has
\begin{align}
&n^2\sigma_{-2}^o(n)= \sigma_2(n) \ ,  \   2  \not|  \  n, \\
&n^2\sigma_{-2}^o(n)=\sigma_{2}(n)-\sum_{2p|n} \frac{n^2}{(2p)^2}=\sigma_2(n)-\sigma_2\left(\frac{n}{2}\right) \ ,  \ 2 \  | \  n  \ . 
\end{align}
This way, one has 
\begin{align}
\sum_{n=1}^{\infty}n^2\sigma_{-2}^{o}(n)z^n=\sum_{n=1}^{\infty}\sigma_2(n)(z^n-z^{2n}) \ . 
\end{align}
One also needs the following
\begin{align}
&\sum_{n=1}^{\infty} \sigma_{k}(n)z^n=\sum_{n=1}^{\infty} \frac{ n^k z^n}{1-z^n } \ . 
\end{align}
The above lead to the relation
\begin{align}
z\frac{\partial}{\partial z}{\cal S}(-z)=4\sum_{n=1}^{\infty}n^2\sigma_{-2}^{o}(n)z^n=4\sum_{n=1}^{\infty}\sigma_2(n)(z^n-z^{2n})=-4\sum_{n=1}^{\infty}\frac{n^2 }{z^{n}-z^{-n}}   \ .
\end{align}
Notice that for $k\in 2\mathbb{Z}_{\ge 1}$, one has the following standard connection to the modular Eisenstein series ~\cite{Stein2003PrincetonLI}
\begin{align}
E_{k}(\tau)=\sum_{\mathbb{Z}^2/(0,0)}\frac{1}{(n\tau+m)^k}=2\zeta(k)+\frac{2(-2\pi \ri)^k}{(k-1)!}\sum_{n\ge 1}\sigma_{k-1}(n)\re^{2\pi n \ri \tau} \ .
\end{align}
In particular, generating functions of odd-power divisor sums can all be expressed in terms of $\vartheta$ function derivatives. However, for odd-$k$, although there is still a similar relation~\cite{gangl2005double}
\begin{align}
\sum_{\substack{n\ge1 \\ m\in \mathbb{Z}}}\frac{1}{(n\tau+m)^k}=\frac{(-2\pi \ri)^k}{(k-1)!}\sum_{n\ge 1} \sigma_{k-1}(n)\re^{2\pi n \ri \tau} \ , 
\end{align}
simple transformation property under $\tau \rightarrow -\frac{1}{\tau}$ no-longer exists, and it is hard to express them in terms of the modular functions. The appearance of the odd-weight non-modular Eisenstein series seems to be a feature that is not shared by the more standard CFT thermal averages.

\section{More details for the main results}
In this section we provide more details for the results in the previous section.  
\subsection{The Mellin-Barnes representation Eq.~(\ref{eq:Barnes})}
We first establish the Barnes representation Eq.~(\ref{eq:Barnes}). Introducing $\kappa=\frac{1}{4N}$ like before, one can write
\begin{align}
\tanh^2 t\kappa=\left(\frac{1-\rho^{2\kappa}}{1+\rho^{2\kappa}}\right)^2 \ , \  \rho=\re^{-t} \ . 
\end{align}
We now Taylor expand the denominator, which is justified since $|\rho^\kappa|<1$
\begin{align}
\frac{1}{(1+\rho^{2\kappa})^2}=1+\sum_{n=1}^{\infty}(-1)^n (n+1)\rho^{2\kappa n} \ . 
\end{align}
We then perform the Mellin-transform on $\kappa$. Due to the presence of the numerator $(1-e^{-2t\kappa})^2$, it is possible to exchange the integration and the summation using the dominated convergence. One needs for $n\ge 1$
\begin{align}
&\int_0^{\infty} \rd\kappa (1-\re^{-2t\kappa} )^2\re^{-2 t n \kappa} \kappa^{u-1}=\frac{\Gamma(u)}{(2t)^u} \left(\frac{1}{n^u}-\frac{2}{(n+1)^u}+\frac{1}{(n+2)^{u}}\right) \  , \\ 
& 0<{\rm Re}(u)<\infty \  , \  n\ge 1 \ , 
\end{align}
and for $n=0$ 
\begin{align}
&\int_0^{\infty} \rd\kappa (1-\re^{-2t\kappa} )^2\kappa^{u-1}=\frac{\Gamma(u)}{(2t)^u} \left(2^{-u}-2\right) \  , \  -2<{\rm Re}(u)<0  \ .  
\end{align}
The upper bound on the convergence strip is $+\infty$ for all the $n\ge 1$ terms due to the exponential decay of $e^{-2tn\kappa}$, while for the $n=0$ term is $0$ due to the convergence requirement at infinity. After Mellin-inversion, one then has the 
\begin{align}
& v\left(\frac{1}{N}\right)=\nonumber \\ 
&-\int_0^{\infty} \rd t \frac{1}{2t(1+\re^{t})}\int_{\rm {\rm Re}(u)>0}\frac{\rd u}{2\pi \ri} (2 \kappa t)^{-u}\Gamma(u)\sum_{n=1}^{\infty}(-1)^n(n+1) \left(\frac{1}{n^u}-\frac{2}{(n+1)^u}+\frac{1}{(n+2)^{u}}\right) \nonumber \\ 
&-\int_0^{\infty} \rd t \frac{1}{2t(1+\re^{t})}\int_{-2<\rm {\rm Re}(u)<0}\frac{\rd u}{2\pi \ri} (2 \kappa t)^{-u}\Gamma(u) (2^{-u}-2) \  . \label{eq:vsum}
\end{align}
Notice that the $n$-sum on the first line converges absolutely, as when ${\rm Re}(u)>0,  \ n\in\mathbb{Z}_{\ge 1}$, one has
\begin{align}
\bigg|\frac{1}{n^u}-\frac{2}{(n+1)^u}+\frac{1}{(n+2)^{u}}\bigg|\le \frac{|u(1+u)|}{n^{2+{\rm Re}(u)}} \ . 
\end{align}
Now, one shifts the contour in the second line of Eq.~(\ref{eq:vsum}) to the ${\rm Re}(u)>0$ region, by adding the negative residue at $u=0$, and combine with the sum over $n\ge 1$, this leads to
\begin{align}
v\left(\frac{1}{N}\right)=-\int_0^{\infty} \frac{\rd t}{2t(1+\re^t)}\bigg(1-4\int_{{\rm Re}(u)>0}\frac{\rd u}{2\pi \ri }\Gamma(u)(4\kappa t)^{-u}(2^u-4)\zeta(u-1)\bigg) \ . 
\end{align}
The term $+1$ is due to the negative residue at $u=0$. Now, to exchange the $\rd t$ and $\rd u$ integrals, we need to shift the $u$-contour to the left again by picking the positive residue at $u=0$, which cancels the $+1$. One then has 
\begin{align}
&v\left(\frac{1}{N}\right)=2\int_{-2<{\rm Re}(u)<0} \frac{\rd u}{2\pi \ri }\Gamma(u)(4\kappa )^{-u}(2^u-4)\zeta(u-1)\int_0^{\infty}\frac{\rd t t^{-u-1}}{1+\re^t} \nonumber \\ 
&=-2\int_{-2<{\rm Re}(u)<0} \frac{\rd u}{2\pi \ri }\Gamma(u)(4\kappa )^{-u}(2^u-4)\zeta(u-1) \times (2^{u+1}-1)\Gamma(-u)\zeta(-u)\ .
\end{align}
This is exactly Eq.~(\ref{eq:Barnes}), after using $\Gamma(u)\Gamma(-u)=-\frac{\pi}{u \sin \pi u }$. 

\subsection{Reflection formulas Eq.~(\ref{eq:reflection1}) and Eq.~(\ref{eq:reflection2})}
We then establish the reflection formulas Eq.~(\ref{eq:reflection1}) and Eq.~(\ref{eq:reflection2}). The idea is to write the $v\left(\frac{1}{N}\right)$ as an infinite sum of log-$\Gamma$ functions, and use the reflection formula for the latter. 

We start again with
\begin{align}
v\left(\frac{1}{N}\right)=-\sum_{n=0}^{\infty}(-1)^n (n+1)\int_0^{\infty}\frac{\rd t}{2t(1+\re^t)} (1-\re^{-2t\kappa})^2\re^{-2tn \kappa} \ . 
\end{align}
Notice that this sum is only conditionally convergent, but when understood as a sum in which the $2k$-th and the $2k+1$-th terms were grouped together, then it is absolutely convergent and there is no issue of exchanging the order of integration and the sum. So for notational simplicity we still write it as a sum over $n$. We will show at the end of this subsection, that the integral above can be performed exactly
\begin{align}
&-\int_0^{\infty}\frac{\rd t}{2t(1+\re^t)} (1-\re^{-2t\kappa})^2\re^{-2tn \kappa} \nonumber \\ 
&=\frac{1}{2}\ln \left(\frac{ \Gamma^2 \left(1+ 2\kappa (n+1)\right)  \Gamma^2 \left(1+\kappa n\right) \Gamma^2 \left(1+\kappa(n+2) \right)}{\Gamma \left(1+2\kappa n\right) \Gamma \left(1+2\kappa (n+2)\right) \Gamma^4 \left(1+\kappa(n+1) \right)}\right) \ . \label{eq:auxi1}
\end{align}
The above should be understood as a sum over the individual $\ln \Gamma(1+\alpha \kappa)$ functions which are positive when $\kappa>0$, and with branch cuts chosen to be along the negative real axis. This way, one has 
\begin{align}
v\left(\frac{1}{N}\right)=\frac{1}{2}\sum_{n=0}^{\infty}(-1)^n (n+1) \ln \left(\frac{ \Gamma^2 \left(1+\frac{  n+1}{2N}\right)  \Gamma^2 \left(1+\frac{ n}{4N}\right) \Gamma^2 \left(1+\frac{n+2}{4N} \right)}{\Gamma \left(1+\frac{n}{2N}\right) \Gamma \left(1+\frac{ n+2}{2N}\right) \Gamma^4 \left(1+\frac{n+1}{4N} \right)}\right) \ .
\end{align}
Using the standard large $z$ expansion of the $\ln \Gamma(1+z)$ function, it is easy to show that the above converges absolutely when even and odd terms are combined together, in the sectors $|{\rm Arg}(\kappa)| \le \pi(1-\delta)$ and defines an analytic function away from the negative real axis. Clearly, this is the same analytic function as defined by Eq.~(\ref{eq:Barnes}), due to the uniqueness of analytic continuation. Given the above, it is not hard to establish the reflection formula Eq.~(\ref{eq:reflection1}), using the standard reflection formula for the $\ln \Gamma(1+z)$ function
\begin{align}
\ln \Gamma(1+z)+\ln \Gamma(1-z)=\ln \frac{\pi z}{\sin \pi z} \ , 
\end{align}
with a branch cut in $z$ along $(-\infty -1)\bigcup(1,\infty)$, but analytic in $(-1,1)$. To simplify further to the Eq.~(\ref{eq:reflection2}), we can write 
\begin{align}
\frac{\tan^2 \pi(n+1)\kappa}{\tan \pi n \kappa \tan\pi(n+2)\kappa}=\bigg(\frac{1-\re^{2\ri\pi n \kappa}}{1+\re^{2\ri\pi n \kappa}}\bigg)\bigg(\frac{1-\re^{2\ri\pi (n+2) \kappa}}{1+\re^{2\ri\pi (n+2) \kappa}}\bigg)\bigg(\frac{1+\re^{2\ri\pi (n+1) \kappa}}{1-\re^{2\ri\pi (n+1) \kappa}}\bigg)^2 \ .
\end{align}
For $\kappa=\frac{1}{4N}=\frac{1}{2}(x+\ri y)$ with $y>0$, one has $|\re^{2\pi i \kappa}|=|\re^{\pi \ri x}e^{-\pi y}|<1$, in this case one can further Taylor expand the logarithms in terms of $e^{\pi i (x+iy)}$, using 
\begin{align}
\ln \frac{1-z}{1+z}=-\sum_{k=0}^{\infty}\frac{2}{2k+1}z^{2k+1} \ , 
\end{align}
and perform the $n$-sum first. This leads exactly to Eq.~(\ref{eq:reflection2}). Again, all the sums are absolutely convergent, if the even and odd terms in the $n$-sum are grouped together. 
 
The integral formula Eq.~(\ref{eq:auxi1}) can be established in various ways. The simplest way is perhaps again the Barnes method. Exactly the same computations in the Eq.~(\ref{eq:vsum}) lead to 
\begin{align}
&-\int_0^{\infty}\frac{\rd t}{t(1+\re^t)} (1-\re^{-2t\kappa})^2\re^{-2tn \kappa} \nonumber \\ 
&=-\int_{-2<{\rm Re}(u)<0} \frac{\rd u}{2\pi \ri}\Gamma(u)(2\kappa)^{-u} \left(\frac{1}{n^u}-\frac{2}{(n+1)^u}+\frac{1}{(n+2)^{u}}\right) \int_0^{\infty} \frac{t^{-u-1}}{1+\re^t}dt \nonumber \\ 
&=-\int_{-2<{\rm Re}(u)<0} \frac{\rd u}{2\pi \ri}\frac{\pi}{u \sin \pi u }(2^{u+1}-1)\zeta(-u)(2\kappa)^{-u} \left(\frac{1}{n^u}-\frac{2}{(n+1)^u}+\frac{1}{(n+2)^{u}}\right) \ . 
\end{align}
Now, unlike the Barnes integral for $v \left(\frac{1}{N}\right)$, the small-$\kappa$ expansion of the above obtained by shifting the contour to the left, has a finite radius of convergence. It can be summed exactly using the Taylor expansion of $\ln \Gamma(1+z)$
\begin{align}
\ln \Gamma(1+z)=-\gamma_Ez +\sum_{l=2}^{\infty}\frac{(-1)^l\zeta_l}{l}z^l \ . 
\end{align}
At the end, all the $\gamma_E$ terms cancel, and it is easy to check that the expansion of the right hand side of Eq.~(\ref{eq:auxi1}) indeed starts from ${\cal O}(\kappa^2)$. In fact, both the first and second Binet's integral and the Hermite's extension can be proven using the Barnes method like here. 

\subsection{The leading singularity Eq.~(\ref{eq:asymdya}); Eq.~(\ref{eq:asym1}) and Eq.~(\ref{eq:sum1})} \label{sec:numberdetails}
Here we prove the leading singularity Eq.~(\ref{eq:asymdya}), the dominant sum Eq.~(\ref{eq:asym1}) and the summation formula Eq.~(\ref{eq:sum1}). 

We start with the summation rule. The idea is to write it as a contour integral 
\begin{align}
\sum_{k=0}^{2n-1} \frac{1}{1+\cos \frac{(2k+1)\pi}{2n}}=\oint_{\cal C} \frac{\rd z}{2\pi \ri z}\frac{-2n}{z^{2n}+1}\frac{1}{1+\frac{1}{2}\left(z+z^{-1}\right)} \ , \ n \in 2 \mathbb{Z}_\ge 1 \ . 
\end{align}
Here, the contour encircles the unit circle counter-clockwisely, but avoids the $z=-1$ double pole. Now, by deforming the contour other way, clearly it can be evaluated as the negative residue at $z=-1$:
\begin{align}
\sum_{k=0}^{2n-1} \frac{1}{1+\cos \frac{(2k+1)\pi}{2n}}=-{\rm Res}(z=-1) \bigg[\frac{1}{z}\frac{-2n}{z^{2n}+1}\frac{1}{1+\frac{1}{2}\left(z+z^{-1}\right)}\bigg]=\frac{1}{2}(2n)^2 \ . 
\end{align}
This leads to Eq.~(\ref{eq:sum1}). 

Now, we establish the dominant sum sum Eq.~(\ref{eq:asym1}). We start with the Eq.~(\ref{eq:realS}). To simplify the notation we write
\begin{align}
c_k=\cos (2k+1)\pi x \ , h_k=\cosh (2k+1)\pi y \ . 
\end{align}
In the region $k\le k_0=[\frac{1}{2\pi|y|}]$, we subtract the $h=1$ version
\begin{align}
\frac{1+c_k h_k}{(c_k+h_k)^2}-\frac{1}{1+c_k}=\frac{(h_k-1)(c_k^2-c_k-h_k-1)}{(c_k+1)(c_k+h_k)^2} \ . 
\end{align}
Now, for $x=\frac{(2q+1)}{2n}$, the Eq.~(\ref{eq:bounds}) is valid, and in the region $k\le k_0 $ one has clearly $|h_k|\le c$ and $|h_k-1| \le c' (2k+1)^2 y^2$. This way, the subtracted sum can be bounded when $|y| \le \epsilon$ as 
\begin{align}
\sum_{k=0}^{k_0}\frac{1}{2k+1}\bigg|\frac{1+c_k h_k}{(c_k+h_k)^2}-\frac{1}{1+c_k}\bigg| \le c''\sum_{k=0}^{k_0} y^2 (2k+1)\le c''' \ , \label{eq:boundsmall}
\end{align}
where $c'''$ is $y$ independent, but depends on $2n$. Then, for the sum in the region $k\ge k_0$, one simply use 
\begin{align}
\bigg|\frac{1+c_k h_k}{(c_k+h_k)^2}\bigg|=\frac{1}{h_k} \bigg|\frac{c_k+\frac{1}{h_k}}{(1+\frac{c_k}{h_k})^2}\bigg|&\le \hat  c \re^{-(2k+1)\pi |y| } \ , \\
\sum_{k=k_0+1}^{\infty}\frac{1}{2k+1}\bigg|\frac{1+c_k h_k}{(c_k+h_k)^2}\bigg|&\le \sum_{k=k_0+1}^{\infty}\frac{ \hat  c}{2k+1}\re^{-(2k+1)\pi|y|} \nonumber \\ 
&< \frac{ \hat  c \re^{-\pi |y|}}{(2k_0+3)(1-\re^{-2\pi |y|})} \le \frac{ \hat  c}{2} \ .  \label{eq:boundlarge}
\end{align}
Combining Eq.~(\ref{eq:boundsmall}) and Eq.~(\ref{eq:boundlarge}) leads to Eq.~(\ref{eq:asym1}). 

Given the above, we can derive now the leading logarithmic singularity Eq.~(\ref{eq:asymdya}) for dyadic rationals. We need to extract the leading small-$y$ singularity of the dominant sum
\begin{align}
{\cal DS}=-\sum_{k=0}^{[\frac{1}{2\pi |y|}]} \frac{2}{2k+1} \frac{1}{1+\cos \frac{(2k+1)(2q+1)\pi}{2^p}} \ . 
\end{align}
For this, we first notice that the  cosine values in the sum are periodic
\begin{align}
&\cos \frac{(2k+1)(2q+1)\pi}{2^p}=\cos \frac{(2k' +1)(2q+1)\pi}{2^p} \ , \\
&k'-k=2^{p}l \ , l\in\mathbb{Z \ . }
\end{align}
This suggests to collect the same cosine values by summing over the $k \mod{2^{p}}$
\begin{align}
{\cal DS}=-\sum_{k=0}^{2^p-1}\frac{1}{1+\cos \frac{(2k+1)(2q+1)\pi}{2^p}}\sum_{l=0}^{[\frac{k_0-k}{2^p}]} \frac{2}{2(k+2^pl)+1} \  , \   k_0=\bigg[\frac{1}{2\pi|y|}\bigg] \ . 
\end{align}
Now, since $k$ takes values in a finite set, when $k_0 \rightarrow \infty$, the leading singularities for each $k$-value can be extracted as 
\begin{align}
\sum_{l=0}^{[\frac{k_0-k}{2^p}]} \frac{2}{2(k+2^pl)+1}=\frac{1}{2^{p}}\ln k_0+{\cal O}(1)=-\frac{1}{2^{p}}\ln |y|+{\cal O}(1) \ . 
\end{align}
Now, after the finite sum over $k$, the ${\cal O}(1)$ terms will not be amplified, and one has
\begin{align}
{\cal DS}=\frac{1}{2^{p}}\sum_{k=0}^{2^p-1}\frac{1}{1+\cos \frac{(2k+1)(2q+1)\pi}{2^p}} \ln |y|+{\cal O}(1)=\frac{2^{p}}{2}\ln |y|+{\cal O}(1) \ . 
\end{align}
For this, we have used the summation rule Eq.~(\ref{eq:sum1}), and the fact that 
the following two sets of cosine values are equal
\begin{align}
\bigg\{\cos\frac{(2k+1)(2q+1)\pi}{2^p} ; \  0\le k\le 2^{p-1}\bigg\} =\bigg\{\cos\frac{(2k+1)\pi}{2^p}; \  0\le k\le 2^{p-1}\bigg\} \ . 
\end{align}
This finishes the derivation of Eq.~(\ref{eq:asymdya}). The leading logarithmic singularities for other rational numbers with factors of $2$ can be established in similar manners.

\subsection{Borel-summability}
Finally, we establish the Borel-summability of the large $N$ expansion. More precisely, we show that the $v\left(\eta=\frac{1}{N}\right)$ when $|{\rm Arg}(\eta)|<\frac{\pi}2{}$ is the Borel re-summation of its small-$z$ or large-$N$ expansion. 

Here we use the Nevanlinna-Sokal theorem~\cite{costin2008asymptotics} to show this. For this, it is sufficient to show that the contour of the Barnes integral Eq.~(\ref{eq:Barnes}), when shifted to the region $-2n-1<{\rm Re}(u)<-2n$, can be bounded by $A^{2n+1}|\eta|^{2n+1} (2n+1)!$ uniformly when $|{\rm Arg}(\eta)|<\frac{\pi}{2}$. After using the functional equation of the $\zeta$ function to change $\zeta(u-1)$ to $\zeta(2-u)$, the remainder to the $2n$-th truncation can be bounded by 
\begin{align}
|R_{2n}(\eta)|\le C_\alpha|2\pi \eta|^{2n+\alpha}\Gamma(2n+2+\alpha)\times\int_{-\infty}^{\infty}\bigg|\frac{\Gamma(2n+2+\alpha+\ri y)}{\Gamma(2n+\alpha+2)(2n+\alpha+\ri y)}\bigg| \rd y \ , \label{eq:remainder}
\end{align}
where we have used the $|{\rm Arg}|(\eta)<\frac{\pi}{2}$ condition, and $C_\alpha$, $0<\alpha<1$ are $n$, $\eta$ independent constants. It remains to show the integral above has an exponential bound. For this, one needs to use the following asymptotics ($1\ll n \rightarrow+\infty$ ):
\begin{align}
&\bigg|\frac{\Gamma(n+\ri \sqrt{n}y)}{\Gamma(n)}\bigg| \nonumber \\ 
&=\exp \bigg(\frac{n-\frac{1}{2}}{2}\ln\left(1+\frac{y^2}{n}\right)-\sqrt{n} y{\rm Arctan}\frac{y}{\sqrt{n}}+O\left(\frac{1}{n}\right)\bigg) \ , 
\end{align}
where the $O \left(\frac{1}{n}\right)$ corrections are uniform in $y$ when $y\in \mathbb{R}$. The above can be established using the standard large argument expansion of the $\ln \Gamma$ function. Using this, it is not hard to show that the integral in Eq.~(\ref{eq:remainder}) has the following large $n$ asymptotics 
\begin{align}
&\int_{-\infty}^{\infty}\bigg|\frac{\Gamma(2n+2+\alpha+\ri y)}{\Gamma(2n+\alpha+2)(2n+\alpha+\ri y)}\bigg| \rd y \rightarrow \frac{\sqrt{2n+2+\alpha}}{2n+\alpha}\int_{-\infty}^{\infty} \rd y \re^{-\frac{y^2}{2}} \ , 
\end{align}
which in fact decays like $n^{-\frac{1}{2}}$. As such, the condition of the Nevanlinna-Sokal theorem is fulfilled and the large $N$ expansion of the $v$ function indeed Borel resums to itself.  Equivalently, we can also use the explicit Borel transform given latter in Eq.~(\ref{eq:Bv}) to directly integrate to the $v\left(\frac{1}{N}\right)$ after changing of variables.

\section{More on the large $N$ factorial asymptotics}
After demonstrating the existence of a natural boundary, in this section, we would like to further locate its origin from the large-$N$ asymptotics. 

For this, it would be helpful to compare with the natural boundaries for thermal averages in finite-volume systems. For example, it is well-known that partition functions and correlation functions for CFTs on torus, sometimes can be written explicitly in terms of the $\vartheta$ functions. These functions depend on the modular parameter $q=\re^{- \frac{2\pi \beta}{L}}$ and are well known to have a natural boundary on the unit circle, corresponding to ${\rm Re}(\tau)={\rm Re}\left(\frac{\beta}{L}\right)=0$, or imaginary temperature/system size. It is not possible to analytic continue up to the negative real axis. The natural boundary in such case is mainly due to the irregularities in the growth speed of the energy-eigenstates numbers, which take discrete values in finite systems and are summed over in the thermal averages.

In our case, we are dealing with a quantity that naively looking would measure ground-state properties in a finite size system, and the thermal average is absent. In the strict CFT limit, such quantities would be given by simple power-laws and clearly free from any natural boundary. In the presence of a finite mass, the analyticity properties of one-point functions in finite-volume QFTs are also likely to be different. For example, the finite-volume one-point function of the spin-operator in the massive Ising~\cite{Fonseca:2001dc} has a finite radius of convergence in the small mass expansion~\cite{Liu:2026qxw}, and the branch points on the imaginary axis seems not singular enough to create natural boundaries (off course, this point needs to be checked further). The physical origin of the natural boundary in our case should be quite different from the known examples in the continuum QFTs. It should be due to certain properties of the Ising-chain at the lattice level, that are lost in the CFT limit.  

\subsection{$\sigma_{-2}(l)$ from factorial asymptotics}    
To further locate such properties, we need to notice that the analytic continuation given by Eq.~(\ref{eq:defv}) is not a randomly chosen one. It is the Borel resummation of the $\frac{1}{N}=\frac{a}{L}$ expansion that measures the ``power-corrections'' to the CFT limit, where $a$ is a ``lattice spacing'' introduced by hand. As such, the natural boundary must lies in the factorial asymptotics of the large-$N$ expansion. To show this, we notice that by shifting the contour to the left in Eq.~(\ref{eq:Barnes}), the large-$N$ expansion reads
    \begin{align}
    v\left(\frac{1}{N}\right)\sim\sum_{n=1}^{\infty}(-1)^n\frac{2 \left(1-\frac{2}{2^{2n}}\right) \left(4-\frac{1}{2^{2n}}\right)  \Gamma (2 n+2)\zeta (2 n) \zeta (2 n+2) }{n (2\pi)^{2n+2}} \left(\frac{1}{N}\right)^{2n}\ .  \label{eq:factorial1}
    \end{align}
    From the above, it is clear that the leading large-$n$ factorial asymptotics that neglects the exponentially-small corrections,
    is given by
 \begin{align}
    v_{\rm samy}\left(\frac{1}{N}\right)\sim\sum_{n=1}^{\infty}(-1)^n\frac{8\Gamma (2 n+2) }{n (2\pi)^{2n+2}} \left(\frac{1}{N}\right)^{2n}\ . 
    \end{align}
    It is again Borel summable. The resulting Borel sum, however, contains only logarithmic branch cuts, but no natural boundaries. 

    This way, we could further narrow the origin of the natural boundary to the exponentially-small corrections multiplying the leading factorial asymptotics. This part is simply given by the factor
 \begin{align}
 t(n)=\left(1-\frac{1}{2^{2n-1}}\right) \left(1-\frac{1}{2^{2n+2}}\right)\zeta(2n)\zeta(2n+2) \ . 
\end{align}
Everything becomes clear after expanding it into different exponential tails (i.e, to the form of a Dirichlet series) using
\begin{align}
&\left(1-\frac{1}{2^{2n-1}}\right)\zeta(2n)\zeta(2n+2)=-\sum_{\substack{p\ge 1\\ q\ge 1}} \frac{(-1)^p}{q^2}\left(\frac{1}{pq}\right)^{2n} \ , \\
&\sum_{q\ge 1}\left(\frac{1}{q}\right)^{2n+2}-\sum_{q\ge 1}\left(\frac{1}{2q}\right)^{2n+2}=\sum_{q \ge 1}\frac{1-(-1)^q}{2}\left(\frac{1}{q}\right)^{2n+2} \ . 
\end{align}
Combining together, one has
\begin{align}
t(n)=\sum_{\substack{p\ge 1\\ q\ge 1}} (-1)^{p-1} \frac{1-(-1)^q}{2} \frac{1}{q^2} \left(\frac{1}{pq}\right)^{2n}=\sum_{l\ge 1}\frac{(-1)^{l-1}}{l^{2n}}\sum_{\substack{q|l\\ q\ge 1}} \frac{1}{q^2}\frac{1-(-1)^q}{2} \ . 
\end{align}
As expected, the divisor sum
\begin{align}
\sum_{\substack{q|l\\ q\ge 1}} \frac{1}{q^2}\frac{1-(-1)^q}{2} \equiv \sigma_{-2}^o(l) \ , 
\end{align}
appears. In terms of $\sigma_{-2}^o(l)$, the large-$N$ asymptotics in Eq.~(\ref{eq:factorial1}) can be written as 
\begin{align}
 v\left(\frac{1}{N}\right)\sim \sum_{n=1}^{\infty}(-1)^n \left(\frac{1}{N}\right)^{2n}\frac{8\Gamma (2 n+2) }{n (2\pi)^{2n+2}} \times \sum_{l=1}^{\infty} \frac{(-1)^{l-1} }{l^{2n}} \sigma_{-2}^o(l) \ , \label{eq:vasymexpanded}
\end{align}
in which the exponentially small corrections attached to the leading factorial speed are made manifest. As such, the origin of the natural boundary has been partly revealed: it is caused by the irregularities in the exponentially-small terms attached to the factorially divergent large-$N$ expansion: for large $l$, $\sigma_{-2}^o(l)$ is known to be bound from the above and below by two constants, but no simple asymptotic formulas can be written for it. 

To fully understand this natural boundary, it remains to see which property of the lattice model could lead to the divisor-sum $\sigma_{-2}^o$ in the large-$N$ expansion of $v\left(\frac{1}{N}\right)$. We have not figured out such properties yet. But it is clear that no natural boundaries could be generated in the one-point functions of the $\epsilon$ and other $\mathbb{Z}_2$-even local operators. The origin of $\sigma_{-2}^o$ should lie in the conversion between the periodic and anti-periodic ground states, and only $\mathbb{Z}_2$-odd operators in the large $N$ expansion could see such irregularities. 

\subsection{Borel singularities}
Since here we have a factorial asymptotics, it is reasonable to comment on its Borel transform and the singularities in the Borel plane.  The Borel transform in our case can be read directly from the integral representation Eq.~(\ref{eq:defv})
\begin{align}
&v\left(\frac{1}{N}\right)=\frac{1}{2}\int_0^{\infty}\rd t \re^{-tN} \sum_{n=1}^{\infty}\frac{(-1)^n\tanh^2\frac{t}{4n}}{t}   \  , \\
& B[v](t)=\frac{1}{2t}\sum_{n=1}^{\infty}(-1)^n\tanh^2\frac{t}{4n} \ . \label{eq:Bv}
\end{align}
It contains infinitely-many double-poles along the imaginary axis at the locations $t=2\pi \ri (2k+1) \times n$. It is not difficult to find the expansion of $B[v]$ at these locations
\begin{align}
B[v](t) \sim -\sum_{l \in \mathbb {Z}/\{0\}}\frac{4\ri (-1)^ll\sigma_{-2}^o(l)}{\pi (t-2\pi \ri l)^2}+\sum_{l \in \mathbb {Z}/\{0\}} \frac{2 (-1)^l\sigma^o_{-2}(l)}{\pi^2(t-2\pi \ri l)}+{\rm regular} \ . 
\end{align}
The divisor sum Eq.~(\ref{eq:defrn}) controls both the single and double poles, and as a consequence, the strengths of the Borel discontinuities across the ${\rm Arg}(t)=\frac{\pi}{2}$ Stokes line
\begin{align}
{\rm disc}_\frac{\pi}{2} v\left(\frac{1}{N}\right)=-\frac{4\ri}{\pi} \sum_{l=1}^{\infty}  (-1)^l\sigma^o_{-2}(l) \left(1+2\pi \ri l N\right) e^{-2\pi \ri lN} \ . 
\end{align}
As such, the presence of a natural boundary is in fact anticipated by the resurgent properties of the large-$N$ asymptotics. 

\subsection{Asymptotics of the leg functions for spin-operator form factors}
Beside the one-point function, the large-$N$ expansion for the spin-operator form-factors~\cite{Liu:2026qxw} themselves can also be worked out explicitly to all orders. In fact, the only quantity that has non-trivial large-$N$ expansion is the following ratios of $\mathfrak{q}$-Pochhammer symbols
\begin{align}
p({\cal Z},\mathfrak{q})=\frac{\prod_{k=0}^{N-1}(\sqrt{{\cal Z}}+\mathfrak{q}^{k})}{\prod_{k=0}^{N-1}(\sqrt{{\cal Z}}+\mathfrak{q}^{k+\frac{1}{2}})} \ ,   \ \mq=e^{\frac{\ri\pi}{N}} \ , \label{eq:anaq}
\end{align}
where $\sqrt{\cal Z}$ is defined in the $(0,2\pi)$ branch. It appears in the one-particle ``leg functions'' for the form factors. In~\cite{Liu:2026qxw}, the above is called a ``square-root product''. To proceed, we follow~\cite{Liu:2026qxw} and introduce the following function 
\begin{align}
\hat \Gamma_N({\cal Z})=-\ri\left(\sqrt{{\cal Z}}-\frac{1}{\sqrt{{\cal Z}}}\right)\int_0^1\frac{\rd t}{2\pi \sqrt{t}}\frac{1+t}{(1-{\cal Z} t)(1-{\cal Z}^{-1}t)}\ln \frac{1-t^N}{1+t^N} \ , 
\end{align}
where the $\sqrt{\cal Z}$ is again in the $(0,2\pi)$ branch. In terms of the $\hat \Gamma_N(\cal Z)$, the product in Eq.~(\ref{eq:anaq}) can be expressed as 
\begin{align}
p({\cal Z},\mq)=\exp \bigg(\hat \Gamma_N({\cal Z})-\frac{1}{2}\ln \left(1-\frac{1}{\sqrt{{\cal Z}}}\right)+\frac{1}{2}\ln \left(1+\frac{1}{\sqrt{{\cal Z}}}\right)\bigg) \ ,  \ {\cal Z} \notin \mathbb{R}_+ \ . 
\end{align}
To work out systematically its large-$N$ expansion, it is convenient to use the following Barnes representation for ${\cal Z} \notin \mathbb{R}_+$~\cite{Liu:2026qxw} 
\begin{align}
\hat \Gamma_N({\cal Z})=-\int \frac{\rd u}{4\ri } \frac{\tan \frac{\pi u}{4N}}{u \cos \frac{\pi u}{2}} (-\ri \sqrt{\cal Z})^{-u},  \  \ -{\rm min} \{1,2N\}<{\rm Re}(u)<{\rm min} \{1,2N\} \ . 
\end{align}
By choosing the contour along the imaginary axis, it is not hard to establish the following integral representation ($0<|k|<N$)
\begin{align}
&\hat \Gamma_N\left(e^{\frac{2\pi \ri k}{N}}\right)-\ln \frac{\sqrt{|k|} \Gamma\left(|k|+\frac{1}{2}\right)}{\Gamma(|k|+1)} =-\int_0^{\infty}\frac{\rd t}{t(1+e^t)}\tanh \frac{t}{4N} \sinh \frac{|k|t}{N} \ . 
\end{align}
From this, the $N\rightarrow \infty$ asymptotics can be worked out explicitly as 
\begin{align}
&\hat \Gamma_N\left(e^{\frac{2\pi \ri k}{N}}\right)-\ln \frac{\sqrt{|k|} \Gamma\left(|k|+\frac{1}{2}\right)}{\Gamma(|k|+1)}\nonumber \\ 
&+ \sum_{n=1}^{\infty}\frac{\left(4^n-2\right)  B_{2 n}  \left(B_{2 n+1}(|k|)+B_{2 n+1}(|k|+1)-2 B_{2 n+1}\left(|k|+\frac{1}{2}\right)\right)}{4 n (2n+1)!} \left(\frac{\ri \pi}{N}\right)^{2n} \ ,  \label{eq:legasym}
\end{align}
where $B_n(x)$ is the standard Bernoulli polynomial. As expected, the expansion diverges factorially. We should  mention that formally one can write the product as
\begin{align}
p(\mathfrak{q}^{2k},\mathfrak{q})=\mathfrak{q}^{-\frac{N}{2}}\frac{(\mathfrak{q}^{k+1};\mathfrak{q})_\infty}{(-\mathfrak{q}^{k+1};\mathfrak{q})_\infty}\frac{(-\mathfrak{q}^{k+\frac{1}{2}};\mathfrak{q})_\infty}{(\mathfrak{q}^{k+\frac{1}{2}};\mathfrak{q})_\infty} \ ,  \ (x;\mathfrak{q})_\infty=\prod_{k=0}^{\infty}(1-x\mathfrak{q}^k) \  ,
\end{align}
and try to use the $\mq \rightarrow 1$ asymptotics in~\cite{Katsurada} for $(\pm \mq^\alpha,\mq)_\infty$. However, since in our case $|\mathfrak{q}|=1$, strictly speaking these asymptotic formulas were not applicable directly. The results above combined with~\cite{Liu:2026qxw} therefore provide an alternative  derivation of the $\mq \rightarrow 1$ asymptotics along the rational points $\mq =e^{\frac{\ri \pi}{N}}$ and $\alpha=k \vee k+\frac{1}{2}$, in a way that avoid the convergence issues. Off course, this method is limited only to situations where the $(x;\mathfrak{q})_\infty$ cancel to finite products $(x;\mathfrak{q})_N$ in ratios.

\section{Comparison with literature}\label{sec:compare}
Finally, in this section we make comparisons with literature and point out certain similarities that might be interesting. The author claims that he is not an expert in string theory/Chern-Simons theory, as such, he can not judge whether the observed similarities are merely superficial coincidences due to the presence of similar $q$-series, or have more explanations. Also, due to the same reason, the literature are no way inclusive. 

First of all, we should mention that factorial asymptotics with two Bernoulli numbers or polynomials, generating divisor-sums in the Dirichlet forms, is not a new mathematical phenomenon. Such asymptotics and their resurgent properties have been studied in literature such as~\cite{Pasquetti:2010bps,Rella:2022bwn,Gu:2023mgf}. Indeed, when expressed in terms of the $B_{2n}$ and $B_{2n+2}$, the factorial asymptotics Eq.~(\ref{eq:factorial1}) takes the following form
 \begin{align}
v\left(\frac{1}{N}\right)\sim4\sum_{n=1}^{\infty}\left(1-\frac{1}{2^{2n-1}}\right)   \left(1-\frac{1}{2^{2n+2}}\right)\left(\frac{2\pi}{N}\right)^{2n}(-1)^{n+1}\frac{ B_{2n+2}}{2n+2} \frac{B_{2n}}{2n}\frac{1}{(2n)!}\ . \label{eq:factorialB}
\end{align}
This is similar in structure to Eq.~(3.32) of~\cite{Gu:2023mgf}. The Dirichlet form Eq.~(\ref{eq:vasymexpanded}) is also similar to Eq.~(5.165) of~\cite{Gu:2023mgf}. On the other hand, the refactor $\left(1-\frac{1}{2^{2n-1}}\right)   \left(1-\frac{1}{2^{2n+2}}\right)$ introduces an additional $(-1)^l$ and projects to the odd-divisors. Also, the asymptotics of the leg function in Eq.~(\ref{eq:legasym}) resembles the asymptotic expansion of the spectral trace ${\rm tr}\rho_{\mathbb{P}^2}$~\cite{Gu:2021ize,Rella:2022bwn}, since both are finite ratios of $(\pm xq^\alpha,q)_\infty$-type symbols. It is reasonable to expect, that natural boundaries possibly with number-theoretical origins are not uncommon and are closely related to the complicated resurgence properties for many examples in~\cite{Pasquetti:2010bps,Garoufalidis:2020xec,Gu:2021ize,Rella:2022bwn,Gu:2023mgf}. 

Moreover, another fact that the author also learned recently~\cite{Marino:2024tbx,Marino:2026}, is that the specific combination of the two Bernoulli numbers  in our asymptotics, the $\frac{ B_{2n+2}}{2n+2} \frac{B_{2n}}{2n}\frac{1}{(2n)!}$, 
is also not new~\cite{Marino:1998pg, faber1998hodgeintegralsgromovwittentheory,Periwal:1993yu,Tierz:2002jj,Marino:2004eq,Marino:2004uf,Gu:2023mgf}. The simplest non-asymptotic explanation that could be made, is that it is possible to express the one-point function directly in terms of the partition function of the Chern-Simons theory on $\mathbb{S}^3$ or of the Stieltjes-Wigert matrix model~\cite{Witten:1988hf,Periwal:1993yu,Marino:2002fk,Tierz:2002jj,Marino:2004uf}.  To show this,  we need to use the fact~\cite{Liu:2026qxw} that the one-point function can be expressed as a finite product of the $p(z,\mq)$ function defined in Eq.~(\ref{eq:anaq})
\begin{align}
&P(\mathfrak{q})=\prod_{k=0}^{N-1} \frac{p\left(\mathfrak{q}^{2k},\mathfrak{q}\right)}{p(\mathfrak{q}^{2k+1},\mathfrak{q})} \ , \label{eq:productP}\\
&G\left(\frac{1}{2}\right)G\left(\frac{3}{2}\right)\left(\frac{2\pi}{N}\right)^{\frac{1}{4}}\re^{v\left(\frac{1}{N}\right)}=\frac{\sqrt{2}}{\sqrt{P(\mathfrak{q})}} \ . 
\end{align}
It is in fact a ratio of two Cauchy determinants made up by the periodic and anti-periodic phases and their square-roots. It is quite interesting to see that the product can be simplified into the integral Eq.~(\ref{eq:defv}). To reach the CS partition function, it is more convenient to use another representation of the leg functions~\cite{Liu:2026qxw}
\begin{align}
&p(\mq^{2k},\mq)=\frac{N}{\mq^k}\frac{\prod_{l}\left(\mq^k-\mq^{l+\frac{1}{2}}\right)}{\prod_{l\ne k}\left(\mq^k-\mq^l\right)} \ ,  \  p(\mq^{2k+1},\mq)=\frac{\mq^{k+\frac{1}{2}}}{N}\frac{\prod_{l\ne k}\left(\mq^{k+\frac{1}{2}}-\mq^{l+\frac{1}{2}}\right)}{\prod_{l}\left(\mq^{k+\frac{1}{2}}-\mq^l\right)} \ . 
\end{align}
Given the above, it is straightforward to simplify the product as 
\begin{align}
&P(\mathfrak{q})=\left(4N^2\sin^2\frac{\pi}{4N}\right)^N \prod_{l=1}^{N-1} \bigg(\frac{\sin\frac{\pi}{2N}\left(l+\frac{1}{2}\right)\sin\frac{\pi}{2N}\left(l-\frac{1}{2}\right)}{\sin^2\frac{\pi l}{2N}}\bigg)^{2N-2l} \nonumber \\ 
&=(4N^2)^N \bigg(\frac{\prod_{l=1}^{2N-1} \left(\sin \frac{\pi l}{4N}\right)^{2N-l}}{\prod_{l=1}^{N-1} \left(\sin \frac{\pi l}{2N}\right)^{4N-4l}}\bigg)^2
\ . 
\end{align}
For comparison, the $U(N)$ CS partition function on $\mathbb{S}^3$ can be written as~\cite{Periwal:1993yu,Tierz:2002jj,Marino:2004uf}
\begin{align}
&Z(N,k)=e^{\frac{\pi \ri}{4}N^2}(k+N)^{-\frac{N}{2}}2^{\frac{N^2-N}{2}}\prod_{l=1}^{N-1}\left(\sin \frac{\pi l }{k+N}\right)^{N-l} \ . 
\end{align}
Combining all above, one has the following representation 
\begin{align}
P(\mq)=\frac{Z^2(2N,2N)}{Z^8(N,N)} \ . 
\end{align}
As such, the large-$N$ expansion can also be obtained from the known perturbative expansion of the CS free energy~\cite{Gopakumar:1998ki,Marino:2004eq,Marino:2004uf,Pasquetti:2010bps} at a fixed $t=\pi \ri $, using 
\begin{align}
g_s= \frac{2\pi \ri }{k+N}\bigg|_{k=N}=\frac{\pi \ri}{N} \ ,  \ t=g_sN= \pi \ri  \ . 
\end{align}
The natural boundary observed here is then along the negative imaginary axis in terms of $g_s$.  In fact, our method provides an alternative derivation of the asymptotic expansion of the CS free energy in the special case $k=N$. For more general $k \in \mathbb{Z}_{\ge 1}$, we expect the method to work as well, after replacing the square-root function to the ${\cal Z}^{\frac{N}{k+N}}$ power-law function. This method leads directly to non-perturbative integral representations like the Eq.~(\ref{eq:defv}), that can also be obtained by Borel re-summing the perturbative expansion~\cite{Pasquetti:2010bps}.

Another comment is for the asymptotics of Lambert series. In~\cite{Banerjee_2017}, the $q \rightarrow 1^-$ asymptotics of a large class of Lambert series was provided. To make a comparison, we write (not to confuse with the $\mathfrak{q}$ in Eq.~(\ref{eq:anaq}) )
\begin{align}
q=\re^{\ri\pi(x+\ri y)}=\re^{2\pi \ri\tau } \ ,  \   \tau= \frac{1}{4N} \ .
\end{align}
Then, the reflection formula for ${\rm Im}(\tau)>0$ reads
\begin{align}
v\left(\tau\right)+v\left(-\tau\right)+\ln \bigg[\pi^{\frac{1}{2}}G^2\left(\frac{1}{2}\right)G^2\left(\frac{3}{2}\right)\bigg] =\frac{1}{2} \ln \frac{\ri}{2\tau}+{\cal S} \left(q\right) \ .
\end{align}
We can now take the $q$-derivatives, using 
\begin{align}
q\frac{\partial}{\partial q}{\cal S}(q)=4{\cal L}_{-q}(2,1)-4{\cal L}_{q^2}(2,1) \ . 
\end{align}
As far as ${\rm Im}(\tau)\ne 0$, the asymptotic expansion in Eq.~(\ref{eq:factorial1}) for $v(\tau)$ is valid. Thus, the above allows also to extract the asymptotic expansion of the Lambert series ${\cal L}_q(2,1)$ as $q \rightarrow -1^+$. 
\begin{figure}[htbp]
    \centering
    \includegraphics[height=6.0cm]{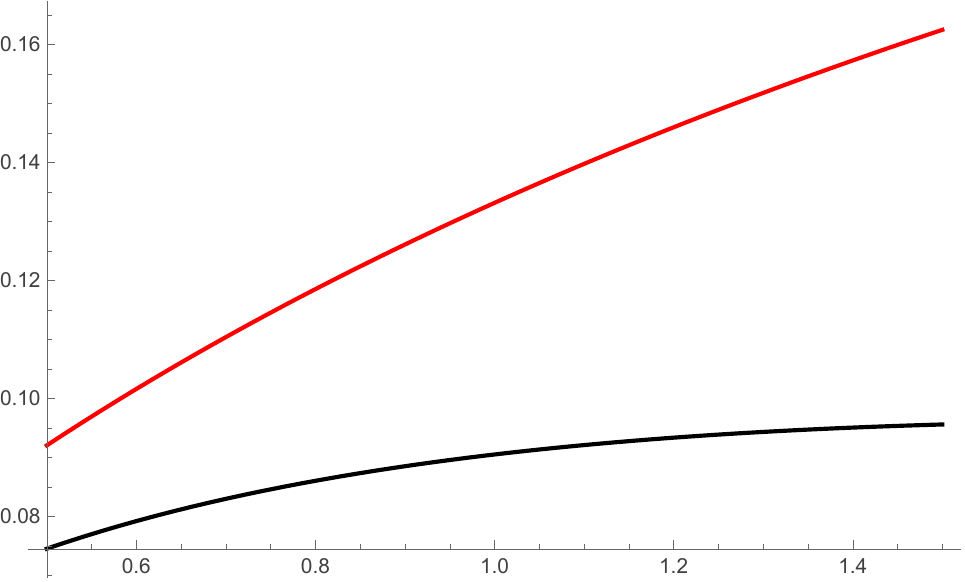}
    \includegraphics[height=6.0cm]{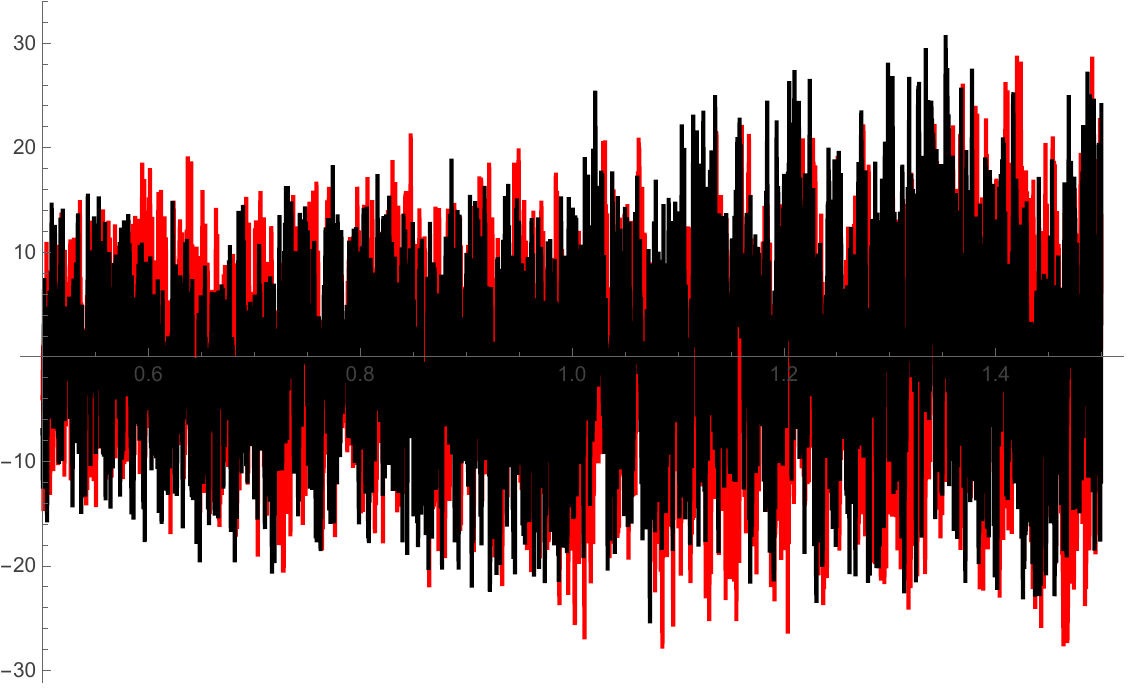}
    \caption{Plots of the right-hand of $J(t+\ri 0)$ in Eq.~(\ref{eq:dual}), for $t=-2\times 10^{-4}\pm\ri y$, with $y\in (0.5,1.5)$. The red and black colors are for real and imaginary parts. The $+$ sign is shown in the upper, while the $-$ sign in shown in the lower. The sums are truncated at $k=800$, which is sufficient for ${\rm Re}(t)=-2\times 10^{-4}$. Clearly, near the ${\rm Arg}(t)=\frac{3\pi}{2}$ axis, the $J(t)$ oscillates extremely fast.}
    \label{fig:Jt}
\end{figure}

Finally, we comment on the analyticity properties of the Mordell-Borel integrals studied in~\cite{Costin:2023kla, Adams:2025aad, Adams:2026ash}. In these literature, they are used to find ``dual'' $q$-series of certain false theta functions across the natural boundary $|q|=1$ along the unit circle, in a way that preserves certain algebraic relations. Here we consider the following integral in the case of $p=3$
\begin{align}
J(t)=\int_0^\infty e^{-\frac{3u^2}{t}}\frac{\sinh u}{\sinh 3u} \rd u \  , \ q=e^{-t} \ ,  \ \tilde q=e^{-\frac{\pi^2}{t}} \ . \label{eq:Moredell}
\end{align}
Clearly, it defines an analytic function in the right-half plane ${\rm Re}(t)>0$. 
Moreover, using the method in~\cite{Costin:2023kla}, it is not hard to establish the following representation
\begin{align}
J(t)=\int_{0<{\rm Re}(u)<1} \frac{\rd u}{4\pi \ri}   \frac{t^{\frac{1-u}{2}}}{2^u(3)^{\frac{1}{2}+\frac{u}{2}}}  \Gamma \left(\frac{1-u}{2}\right) \Gamma (u)\left(\zeta \left(u,\frac{1}{3}\right)-\zeta \left(u,\frac{2}{3}\right)\right) \ . 
\end{align}
From the above, it is clear that the analyticity region is at least
\begin{align}
-\frac{3\pi}{2}<{\rm Arg}(t)<\frac{3\pi}{2} \ . 
\end{align}
However, when analytic continued to the threshold value  ${\rm Arg}(t)=\pm \frac{3\pi}{2}$, like the Barnes representation Eq.~(\ref{eq:Barnes}), the exponential decay is canceled and there are real singularities. These singularities are not related to the $|q| \rightarrow 1^-$ singularities of mock theta functions in the $t>0$ decomposition. Instead, to show that one can not go beyond $\pm \frac{3\pi}{2}$, one needs the $t<0$ decomposition in terms of false theta functions~\cite{Adams:2025aad}
\begin{align}
J(t\pm \ri 0)= \mp \ri t\sqrt{\frac{\pi}{-12t}}\Psi^2_3\left(\frac{1}{q}\right)+\frac{\pi}{\sqrt{12}}\Psi^2_3\left(\frac{1}{\tilde q }\right) \ , \ t\in \mathbb{R}_- \ , \  {\rm Arg}(t\pm \ri 0)=\pm \pi \ ,  \label{eq:dual}
\end{align}
where one has
\begin{align}
\Psi^2_3\left(\frac{1}{q}\right)=\sum_{k=0}^{\infty}\left(e^{t \frac{(3k+1)^2}{3}}-e^{t \frac{(3k+2)^2}{3}} \right) \ . 
\end{align}
Let's consider the upper branch $J(t+\ri 0)$. The decomposition Eq.~(\ref{eq:dual}) allows its analytic continuation to the entire left half-plane, corresponds exactly to the range ${\rm Arg}(t)\in [\frac{\pi}{2},\frac{3\pi}{2})$ for $J(t)$. Along the positive imaginary axis, it is regular. However, when approaching the negative imaginary axis, the singularities no longer cancel and one reaches the natural boundary of $J(t)$ at ${\rm Arg}(t)=\frac{3\pi}{2}$. 
To see this, it is convenient to introduce $t=x+\ri y$ and separate the corresponding real and imaginary parts through
\begin{align}
&- t\sqrt{\frac{\pi}{-12t}}\Psi^2_3\left(\frac{1}{q}\right)=A(x,y)+\ri B(x,y) \ , \\
&\frac{\pi}{\sqrt{12}}\Psi^2_3\left(\frac{1}{\tilde q }\right)=\tilde A(x,y)+\ri \tilde B(x,y) \ .
\end{align}
Using the fact that $A(x,y)$, $ \tilde A(x,y)$ are even in $y$, while $B(x,y)$, $\tilde B (x,y)$ are odd in $y$, for $\frac{\pi}{2}<{\rm Arg}(t)<\frac{3\pi}{2} $, one has
\begin{align}
J(x,y)+J(x,-y)=2\tilde A(x,y)+2\ri A(x,y) \ . \label{eq:reflectionJ}
\end{align}
As such, when $x\rightarrow 0^-$, the real and imaginary parts of the even-$y$ projection are each given by a single $q$-series, for which no cancellation is possible. Thus, since for $y>0$, $J(x,y)$ has no natural boundary when $x\rightarrow 0^-$, there must be one at $y<0$. This asymmetry is due to the presence of an $\ri$ in the decomposition Eq.~(\ref{eq:dual}). In this sense, the ${\rm Arg}(t)=\pi \mp \frac{\pi}{2}$ are similar to the positive and negative real axis in our case, and the relation Eq.~(\ref{eq:reflectionJ}) plays a similar role as our reflection formula Eq.~(\ref{eq:reflection2}), for the purpose of establishing a natural boundary. Notice that in this example, the false theta functions in the $t<0$ decomposition that encode the natural boundary can be recovered from resurgence analysis~\cite{Adams:2025aad} for the $t\rightarrow0$ and $t\rightarrow \infty$ asymptotics. In Fig.~\ref{fig:Jt}, we show the behavior of the right-hand side of Eq.~(\ref{eq:dual}) for $J(t+\ri 0)$ near the positive and negative imaginary axis at ${\rm Re}(t)=-2\times 10^{-4}$, to convince the readers that a natural boundary at ${\rm Arg}(t)=\frac{3\pi}{2}$ exists.

\bibliographystyle{apsrev4-1} 
\bibliography{bibliography}

\end{document}